\documentclass[aps,superscriptaddress,amsmath,amssymb,twocolumn,amsfonts,longbibliography]{revtex4-1}
\usepackage{graphicx}
\usepackage{hyperref}
\usepackage{cleveref}
\usepackage{color}
\usepackage{float}
\usepackage{soul}
\usepackage{ulem}
\usepackage{braket}
\usepackage{dsfont}
\usepackage{physics}
\usepackage{tensor,upgreek,float}
\usepackage{parskip,tipa}
\usepackage{tikz}

\newcommand\source[1]{%
    \tikz[remember picture,baseline,inner sep=0pt] {%
        \node [name=source,anchor=base]{$#1$};
    }%
    \setcounter{target}{0}
}
\newcounter{target}
\newcommand\target[1]{%
    \tikz[remember picture,baseline,inner sep=0pt] {%
        \node [name=target-\thetarget,anchor=base]{$#1$};
    }%
    \stepcounter{target}%
}
\newcommand\drawarrows{
    \tikz[remember picture, overlay, bend left=45, -latex] {
        \foreach \i [evaluate=\i as \n using int(\i-1)] in {1,...,\thetarget} {
            \draw ([xshift=-1mm,yshift=0.5mm]source.north) to ([xshift=1.5mm,yshift=0.5mm]target-\n.north);
        }
    }
}

\newcommand{\uk}[1]{\textcolor{red}{#1}}

\begin{document}
\author{Prabhakar }
\affiliation{School of Physical Sciences, National Institute of Science Education and Research, a CI of Homi Bhabha National Institute, Jatni 752050, India}
\title{Predicting Fractionalized Multi-Spin Excitations in Resonant Inelastic X-ray Spectra of Frustrated Spin-1/2 Trimer Chains}

\author{Subhajyoti Pal}
\affiliation{School of Physical Sciences, National Institute of Science Education and Research, a CI of Homi Bhabha National Institute, Jatni 752050, India}

\author{Umesh Kumar}
\affiliation{Department of Physics and Astronomy, Rutgers University, Piscataway, NJ 08854, USA}

\author{Manoranjan Kumar}
\affiliation{S. N. Bose National Centre for Basic Sciences, Block JD, Sector III, Salt Lake, Kolkata 700106, India}

\author{Anamitra Mukherjee}
\email{anamitra@niser.ac.in}
\affiliation{School of Physical Sciences, National Institute of Science Education and Research, a CI of Homi Bhabha National Institute, Jatni 752050, India}

\date{\today}

\date{\today}
\begin{abstract}
We theoretically investigate the resonant inelastic X-ray scattering (RIXS) spectra in a quasi-1D chain of weakly coupled frustrated spin-1/2 trimers, as realized in Na$_{2}$Cu$_{3}$Ge$_{4}$O$_{12}$, with  Cu $d^{9}$ 1/2 spins. We compute multi-spin correlations contributing to spin-conserving (SC) and spin non-conserving (NSC) RIXS cross-sections using ultra-short core-hole lifetime expansion within the Kramer-Heisenberg formalism. These excitations involve flipping spins of up to three spin-1/2 trimers and include the inelastic neutron scattering (INS) single spin-flip excitations in the lowest order of the NSC channel. 
We identify the fractionalization 
of two coupled frustrated trimers in terms of spinons, doublons, and quartons in the spectra evaluated using exact diagonalization, complementing prior studies single spin-spin flip excitation in inelastic neutron scattering. 
Specifically, we uncover two new high-energy modes at $\omega \approx 2.4J_1$ and $3.0 J_1$ in the NSC and SC channels that 
are accessible at the Cu $K$-edge and $L$-edge RIXS spectra, which were missing in the INS study. This, therefore, provides pathways to uncover all the possible excitations in coupled trimers. Our work opens new opportunities for understanding the nature of fractionalization and RIXS spectra of frustrated, low-dimensional spin chains.
\end{abstract}

\maketitle

\section{Introduction}
\label{intro}
\begin{figure}[t]
    \centering{
    \includegraphics[width=1\linewidth]{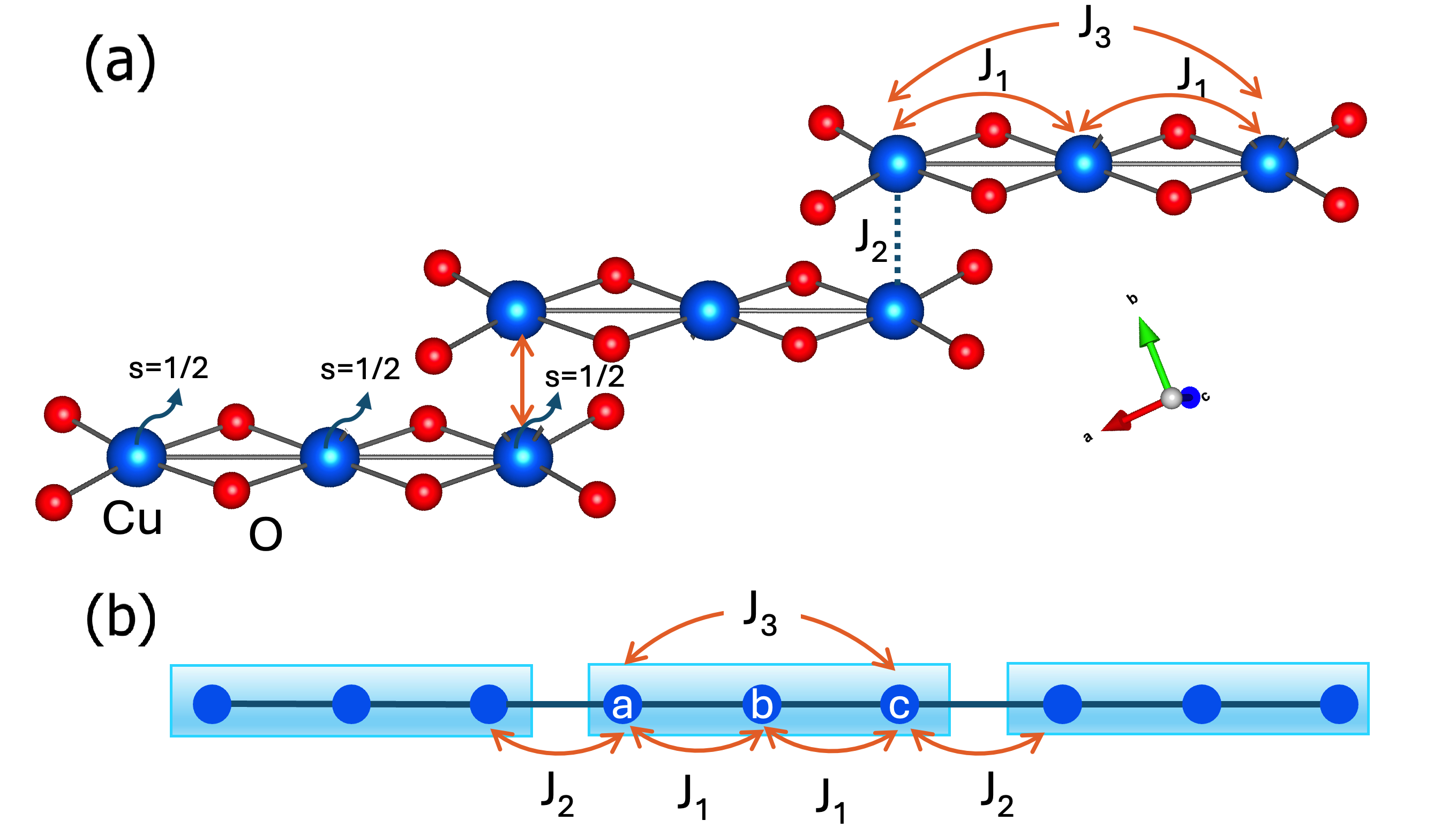}}
    \caption{(a) Shows the schematics of [Cu$_3$O$_8$]$^{-10}$ spin-1/2 trimers in Na$_{2}$Cu$_{3}$Ge$_{4}$O$_{12}$, The magnetic exchanges within the trimer are $J_1$ and $J_3$, while $J_2$, couples two trimers. The schematic representation of one dimensional spin-1/2 trimer chain extracted from Na$_{2}$Cu$_{3}$Ge$_{4}$O$_{12}$ is shown in (b). The blue circles represent the Cu ions hosting spin 1/2 magnetic moments. $J_{1}$, $J_{2}$, and $J_{3}$ correspond to the nearest-neighbor intra-trimer, nearest-neighbor inter-trimer, and next-nearest-neighbor intra-trimer interactions. The rectangle around the trimers contains three Cu atoms labeled `a', `b', and `c' and will be useful in RIXS calculations later. We note that within each trimer, $J_3$ acts as a source of frustration.}
    \label{fig:lattice}
\end{figure}

Since the breakthrough discovery of high-temperature superconductivity in cuprates \cite{RevModPhys.66.763}, cuprates have continued to be at the forefront of research in condensed matter physics. These materials have undergone extensive experimental and theoretical investigation, yet they continue to yield surprising behavior. Alongside their manifestation in 2D geometries \cite{2dcuprate1,2dcuprate2,2dcuprate3,2dcuprate4}, cuprates manifest in one-dimensional (1D) chain \cite{1dcuprate1,PhysRevLett.77.4054} and ladder structures \cite{PhysRevB.45.5744}. The 1D geometries exhibit distinctive behaviors and serve as a platform for exploring many-body physics. Magnetism, in one dimension, is strongly influenced by quantum fluctuations and has been thoroughly investigated since the beginning of quantum mechanics. The Bethe ansatz \cite{Bethe1931} based solution of the 1D Heisenberg antiferromagnetic  chain (HAC) is a cornerstone result and has deepened our understanding of low dimensional interacting systems \cite{BA1,BA2}. However, it is well-known that the integrability of the HAC is destroyed once longer range spin-interactions are introduced. One such system of interest is the spin-1/2 trimer chain \cite{prb_trimer,trimer_jpsj,Cheng2022}, a 1D arrangement of three spin unit cells that can exhibit unique quantum phenomena. These trimer chains are known to exhibit intriguing properties such as fractionalized excitations and exotic magnetic states. 
 \textcolor{black}{Recently, a new family of cuprates X$_2$Cu$_3$Ge$_{4-x}$Si$_x$O$_{12}$ with X=Li, Na has been discovered\cite{,trimer_exp2}. The x=0 compound consists of  [Cu$_3$O$_8$]$^{-10}$ spin-trimers, with each Cu acting as a spin S=1/2. These trimers are periodically arranged as shown in Fig.~\ref{fig:lattice} (a) and are coupled by magnetic exchanges in addition to magnetic exchanges coupling spins within the trimer. Continued experimental effort \cite{trimer_exp}, and detailed materials theoretic modeling has revealed an effective model of frustrated trimers coupled with magnetic exchanges shown in Fig.~\ref{fig:lattice} (b).}
 
A recent study of inelastic neutron scattering  (INS) experiment on Na$_{2}$Cu$_{3}$Ge$_{4}$O$_{12}$  revealed signature of fractional excitation termed as spinon, doublon and quarton modes \cite{Bera2022}. However, higher-order spin excitations have yet to be reported due to the limitations in the energy range of the INS. 
Apart from new high energy features, current theoretical analysis has that have so far reported single spin-flip excitations only. The contributions of multi-spin excitations in the energy range of dynamical (single-spin-flip) excitations have remained open for frustrated spin-trimer chains. The latter is important in constraining sum-rules, as has been done for HAC recently~\cite{Mourigal2013}. Resonant Inelastic X-ray Scattering (RIXS) has emerged as a standard probe for measuring multi-spin excitations \cite{RevModPhys.83.705,van_den_Brink_2007,J.van_den_Brink_2006,Peng2017,PhysRevLett.104.077002,PhysRevLett.103.117003}. We thus focus of the theory of RIXS-inspired computation of multi-spin correlation functions. 

Resonant Inelastic X-ray Scattering (RIXS)~\cite{RevModPhys.83.705,RevModPhys.93.035001, deGroot2024, van_den_Brink_2007,J.van_den_Brink_2006,Peng2017,PhysRevLett.104.077002,PhysRevLett.103.117003} has become an invaluable, powerful spectroscopic tool for probing complex materials' electronic structure, magnetic properties, and lattice dynamics. By tuning the incident X-ray energy to an absorption edge, RIXS provides element-specific orbital-selective insights into various excitations, ranging from charge transfer and spin waves to phonon and orbital excitations. This makes it particularly useful for studying low-dimensional systems where interactions between electronic, magnetic, and lattice degrees of freedom are often pronounced. RIXS has enabled the detection of multi-spinon excitation \cite{multi_spinon1,Schlappa2018}, spin-orbital separation \cite{spin_orb_sepr, PhysRevB.88.195138} in antiferromagnetic Heisenberg material Sr$_2$CuO$_{3}$ and high energy spin excitations in the quantum spin liquid candidate Zn-substituted barlowite \cite{PhysRevB.107.L060402}. In recent years, the improved resolution in the meV scale has opened new opportunities to explore magnetism in materials such as Na$_{2}$Cu$_{3}$Ge$_{4}$O$_{12}$~\cite{Zhou:rv5159}.
The Kramers-Heisenberg (KH) formalism \cite{RevModPhys.83.705}, used to simulate the RIXS cross-section, is complex and makes interpreting RIXS data difficult. Nevertheless, significant progress has been made in exploring quantum magnets with RIXS. The ultrashort core-hole lifetime (UCL) expansion in RIXS expands the cross-section into spin-conserving (SC) and non-spin-conserving (NSC) channels for the $K$ and $L$-edge RIXS respectively~\cite{UKumar2022, PhysRevX.12.021041, PhysRevX.6.021020, UKumar2019, PhysRevB.77.134428, PhysRevB.85.064421}.  
For the cuprates, usually at $L$-edge, $J_1/\Gamma < 1$, where $\Gamma$ is the inverse of the core-hole life time. Due to this, one can expand the RIXS cross-section in orders of $1/\Gamma$. Due to richer statistics in RIXS, higher-order corrections can be experimentally observed~\cite{UKumar2022}. The  NSC channel of RIXS is typically considered analogous to the INS probe for the lowest order of the UCL expansion, with higher-order contributions remaining largely unexplored in the literature. On the other hand, the SC channel successfully determines the four spinons \cite{UKumar2018,Schlappa2018} in 1D cuprates and multi-magnons \cite{PhysRevB.103.224427,PhysRevB.108.214405}, multi-triplons \cite{PhysRevLett.98.027403,PhysRevLett.87.127002} in 2D cuprates which are inaccessible in INS experiments.

In this work, we investigate resonant inelastic X-ray scattering (RIXS) in quasi-1D spin-1/2 trimer chains, focusing on  $\text{Na}_{2}\text{Cu}_{3}\text{Ge}_{4}\text{O}_{12}$. Through Lanczos diagonalization and Fock-space Recursive Green's function (F-RGF) technique, we uncover a richer fractionalization pattern than previously reported. We identify two new high-energy modes in both the NSC and SC channels. We provide a detailed understanding of the single and the multiple spin-flip processes in terms of spinons, doublons, and quartons. We also report weakly dispersing fractionalized modes in multi-spin correlations. 

The paper is structured as follows. In Sec. \ref{model}, we first introduce the spin model for the 1D antiferromagnetic Heisenberg trimer chain investigated in this paper. We then explain the response functions used for the corrections in the UCL expansion of the RIXS cross-section in SC as well as NSC channels and provide the expression for various operators in detail. We present our results in Sec. \ref{results} and conclude the paper in Sec. \ref{concl}.

\section{Hamiltonian \& RIXS cross-section}
\label{model}
\subsection{Hamiltonian}
 $\text{Na}_{2}\text{Cu}_{3}\text{Ge}_{4}\text{O}_{12}$ is known to be the host of the frustrated spin-half antiferromagnetic Heisenberg chain of trimers \cite{trimer_exp}. The effective spin Hamiltonian\cite{Bera2022} shown in Fig.~\ref{fig:lattice} (b) is given by,
\begin{equation}
\begin{split}
H=&\sum_{r=0}^{N/3-1}[J_{1}(\vec{S}_{r,a}\cdot \vec{S}_{r,b}+\vec{S}_{r,b}\cdot \vec{S}_{r,c})\\
&+J_{2}\vec{S}_{r,c}\cdot \vec{S}_{r+1,a}+J_{3}\vec{S}_{r,a}\cdot \vec{S}_{r,c}].
\end{split}
\label{eq:ham}
\end{equation}
The unit cell consists of three spins, as shown in Fig.~\ref{fig:lattice}, and summation $r$ is over the unit cell. Where $\vec{S}_{r,\mu}$ is the spin operator at $\mu ^{th}$(=a,b,c) site of $r^{th}$ trimer. $J_{1}$,$J_{2}$$(=\alpha J_{1})$ and $J_{3}$$(=\beta J_{1})$ are the nearest-neighbor intra trimer, nearest-neighbor inter trimer and next-nearest-neighbor intra-trimer interaction respectively. $\beta$ is a frustration parameter for isolated trimers. From the materials analysis it is known that both $\alpha$ and $\beta$ are equal to 0.18 \cite{Bera2022}. Thus, for the present material, we are at the limit of weak frustration and weak inter-trimer coupling.

 \begin{figure}[t]
	\centering
	\includegraphics[width=1.0              \linewidth]{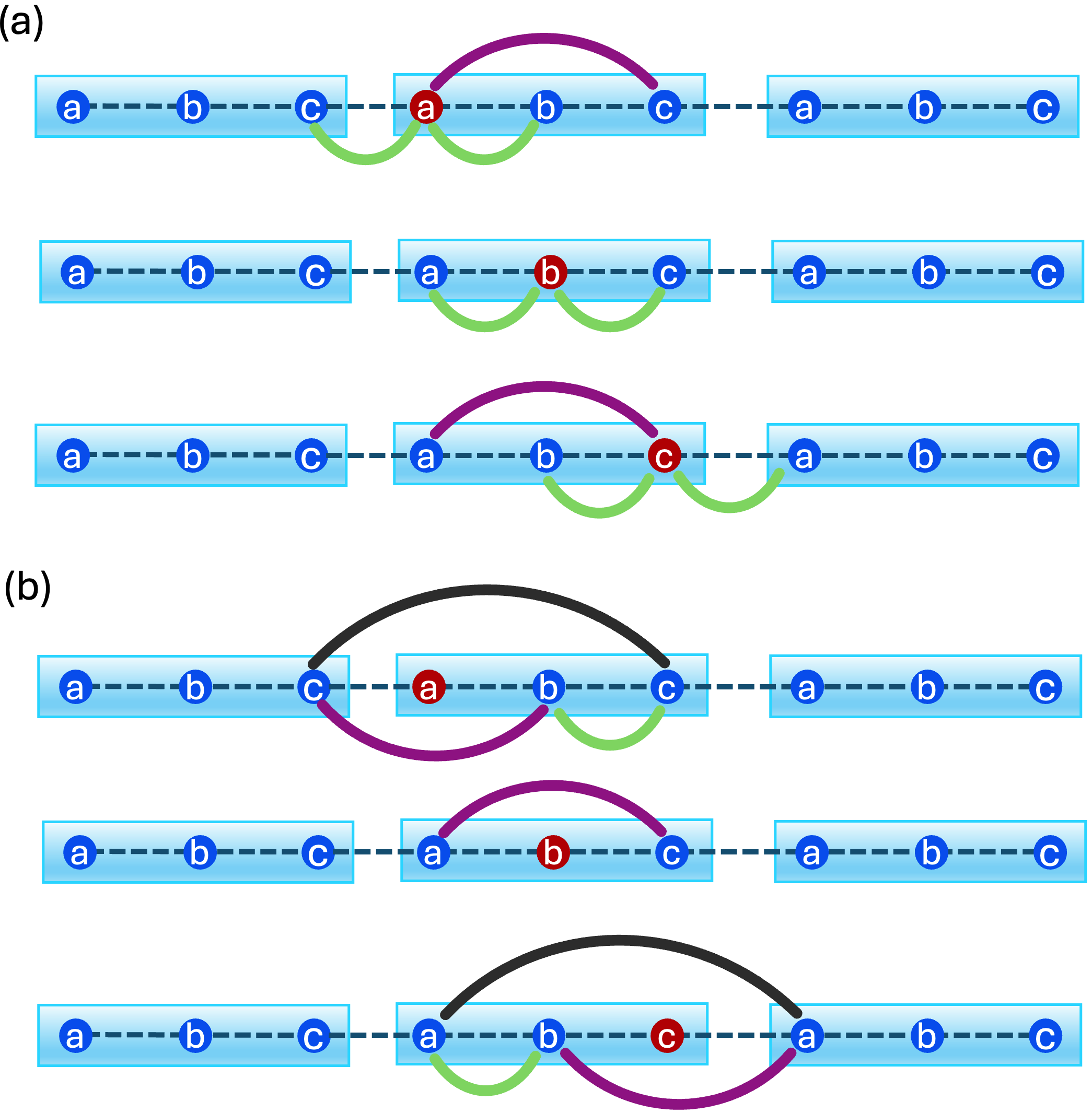}
\caption{Connectivity of the RIXS operators $O_{i,l}^{NSC/SC}$: Core hole creation site $i$ is marked in red. Panel (a) illustrates the connectivity for the first order in UCL expansion in both the spin-conserving (SC) and non-spin-conserving (NSC) channels. Panel (b) depicts the connectivity in second order for both the spin-conserving (SC) and non-spin-conserving (NSC) channels. We note that the connectivity of the SC and NSC channels are identical at any given order; the essential difference comes from the spin-flip at the core-hole creation site in the NSC channel.}
\label{fig:operator_coonec}
\end{figure}
\subsection{RIXS cross-section}
RIXS is a photon-in photon-out process. The RIXS process measures the scattering cross-section of an incoming X-ray whose energy is tuned to the energy difference between specific core atomic energy level and valence band of a material under consideration. For transition metal oxides, the incoming  X-ray energy can be made resonant with the core levels of either the transition metal atom or Oxygen in a transition metal oxide. Depending on incident energy, the X-ray can cause a photo-excitation from  $1s$ to $4p$ ($K$-edge) or $2p$ to $3d$ ($L$-edge). It has been well-established that due to the spin-orbit interaction in the 2p state, the $L$-edge can probe spin angular momentum non-conserving spin excitation processes \cite{PhysRevB.77.134428, PhysRevLett.103.117003}, while $K$-edge probes only spin angular momentum conserving spin-excitations. \textcolor{black}{We note that the $L$-edge can also probe spin-conserving excitations as well.}
For an incoming photon with momentum $q_\text{in}$, and an outgoing photon with momentum $q_\text{out}$, the experiment involves measuring the intensity of the scattered radiation as a function of energy loss and momentum transfer. 

\textcolor{black}{In the present work, we envisage a RIXS ($L$-edge) experiment at the Cu atoms in the spin-trimer that can probe both the NSC and SC channels.} Hence, we investigate the spin Hamiltonian given by Eq.~\eqref{eq:ham} within the UCL approximation of the Kramer-Heisenberg (KH) formalism. RIXS intensity decomposed into correlation functions in spin-conserving (SC) and non-spin-conserving (NSC) channels in UCL expansion. The RIXS cross-section is given by KH formalism; $I_\text{RIXS} = \sum_{f}\vert \langle f| D_\text{out} \mathcal{O} D_\text{in}|g\rangle|^2 \delta (E_f-E_g-\omega)$. Here, $|g\rangle(|f\rangle)$ are the ground (final) states from the Hamiltonian $H_0$ with energy $E_g~(E_f)$. $q(= q_\text{out} - q_\text{in})$ and $\omega$ is the momentum transfer and  the energy loss respectively. $D_\text{in(out)}$ is the dipole operator and $\mathcal{O}$ accounts for the evolution of the system in the presence of the core hole. We refer to recent literature for a detailed exposition of the simplification of the cross-section (See Ref.~\cite{UKumar2022}). Following the literature~\cite{UKumar2022, PhysRevX.6.021020}, we employ ultra-short core-hole lifetime approximation. This introduces a broadening factor ($\Gamma$), which is the inverse of the core-hole lifetime. The RIXS intensity $I_\text{RIXS} \propto \sum_{l} \chi_{l}^\text{NSC}(q,\omega)+\sum_{l}\chi_{l}^\text{SC}(q,\omega)$, the label `$l$' refers to the order of expansion in the inverse of the core-hole lifetime, as discussed below. The proportionality constant involves polarization-dependent matrix elements emerging from the dipole operators and differs for the NSC and SC channels \cite{PhysRevLett.103.117003}. We provide the detailed derivation RIXS cross-section in Appendix \ref{appendix}.

\subsubsection{Non spin conserving (NSC) channel}
The $l^{th}$ order terms in the UCL expansion in the non-spin-conserving channel is,

\begin{equation}
\begin{split}
\chi_{l}^{NSC}(q,\omega)=\frac{1}{\Gamma^{2l+2}}&\sum_{f}\left\vert\langle f| \frac{1}{\sqrt{N}}\sum_{i=1}^{N} e^{\iota q r_{i}}O_{i,l}^{NSC} |g\rangle \right\vert^2 \\
&\times\delta\left(E_{f}-E_{g}- \omega\right)
\end{split}
\label{eq:rixs_nsc}
\end{equation}

Where $\ket{g}$ ($\ket{f}$) is the ground (final) state with energy $E_{g}$ ($E_{f}$) of $H$, and summation $i$ is over the sites. 
Schematics of connectivity in the various RIXS scattering operators $O_{i,l}^{NSC/SC}$ are shown in Fig.~\ref{fig:operator_coonec}. Panel (a) (panel (b)) depicts the connectivity for the first order (second order) in the NSC and SC channels. The connectivity of the RIXS scattering operators arising from the UCL expansion for the NSC and SC channels are identical. The difference between the processes originates from spin-flip at the core-hole creation site (marked in red in Fig.~\ref{fig:operator_coonec}) in the NSC channel. 
\\

\textit{\underline{Zeroth order:}} $0^{th}$ order term in NSC channel,

\begin{equation} 
 O^{NSC}_{i,0}=S^{x}_{i}
 \label{eq:nsc0}
\end{equation}
 
\textit{\underline{First order:}} $1^{st}$ order correction in NSC channel,

if $i\in a$;
\begin{equation} 
 \begin{split}
 O^{NSC}_{i,1}=&J_{1} {S}^{x}_{i}(\vec{S}_{i}\cdot \vec{S}_{i+1})+J_{2} {S}^{x}_{i}(\vec{S}_{i}\cdot \vec{S}_{i-1})\\
& +J_{3} {S}^{x}_{i}(\vec{S}_{i}\cdot \vec{S}_{i+2})
\end{split}
\label{eq:nsc1_a}
\end{equation}
if $i\in b$;
\begin{equation} 
 O^{NSC}_{i,1}=J_{1} {S}^{x}_{i}(\vec{S}_{i}\cdot \vec{S}_{i+1})+J_{1} {S}^{x}_{i}(\vec{S}_{i}\cdot \vec{S}_{i-1})
 \label{eq:nsc1_b}
\end{equation}
if $i\in c$;
\begin{equation} 
 \begin{split}
 O^{NSC}_{i,1}=&J_{1} {S}^{x}_{i}(\vec{S}_{i}\cdot \vec{S}_{i-1})+J_{2} {S}^{x}_{i}(\vec{S}_{i}\cdot \vec{S}_{i+1})\\
 &+J_{3} {S}^{x}_{i}(\vec{S}_{i}\cdot \vec{S}_{i-2})\end{split}
 \label{eq:nsc1_c}
\end{equation}

\textit{\underline{Second order:}} $2^{nd}$ order correction in NSC channel,

if $i\in a$;
\begin{equation} 
 \begin{split}
 O^{NSC}_{i,2}=&J_{1}J_{2} {S}^{x}_{i}(\vec{S}_{i}\cdot \vec{S}_{i-1})(\vec{S}_{i}\cdot \vec{S}_{i+1})
 +J_{1}J_{3} {S}^{x}_{i}(\vec{S}_{i}\cdot \vec{S}_{i+1})\\
 &(\vec{S}_{i}\cdot \vec{S}_{i+2})
+J_{2}J_{3} {S}^{x}_{i}(\vec{S}_{i}\cdot \vec{S}_{i-1})(\vec{S}_{i}\cdot \vec{S}_{i+2})
\end{split}
\label{eq:nsc2_a}
\end{equation}
if $i\in b$;
\begin{equation} 
     O^{NSC}_{i,2}=J_{1}J_{1} {S}^{x}_{i}(\vec{S}_{i}\cdot \vec{S}_{i+1})(\vec{S}_{i}\cdot \vec{S}_{i-1})
     \label{eq:nsc2_b}
\end{equation}
if $i\in c$;
\begin{equation} 
 \begin{split}
 O^{NSC}_{i,2}=&J_{1}J_{2} {S}^{x}_{i}(\vec{S}_{i}\cdot \vec{S}_{i+1})(\vec{S}_{i}\cdot \vec{S}_{i-1})+J_{1}J_{3} {S}^{x}_{i}(\vec{S}_{i}\cdot \vec{S}_{i-1})\\
 &(\vec{S}_{i}\cdot \vec{S}_{i-2})
+J_{2}J_{3} {S}^{x}_{i}(\vec{S}_{i}\cdot \vec{S}_{i+1})(\vec{S}_{i}\cdot \vec{S}_{i-2})
\end{split}
\label{eq:nsc2_c}
\end{equation}

 \subsubsection{Spin conserving (SC) channel}
 The $l^{th}$ order terms in the UCL expansion in the spin-conserving channel
is,
\begin{equation}
\begin{split}
\chi_{l}^{SC}(q,\omega)=\frac{1}{\Gamma^{2l+2}}&\sum_{f}\left\vert\langle f| \frac{1}{\sqrt{N}}\sum_{i=1}^{N} e^{\iota q r_{i}}O_{i,l}^{SC} |g\rangle \right\vert^2 \\
&\times\delta\left(E_{f}-E_{g}- \omega\right)
\end{split}
\label{eq:spectra_sc}
\end{equation}

Where $\ket{g}$ ($\ket{f}$) is the ground (final) state with energy $E_{g}$ ($E_{f}$) of $H$ and summation $i$ is over the sites. Operators belonging to each order are written explicitly below. \\
\\
\textit{\underline{First order:}}

if $i\in a$;
\begin{equation} 
\begin{split}
 O^{SC}_{i,1}=&J_{1}(\vec{S}_{i}\cdot \vec{S}_{i+1})+J_{2}(\vec{S}_{i}\cdot \vec{S}_{i-1})\\
& +J_{3} (\vec{S}_{i}\cdot \vec{S}_{i+2})
\end{split}
\label{eq:sc1_a}
\end{equation}
if $i\in b$;
\begin{equation} 
O^{SC}_{i,1}=J_{1}(\vec{S}_{i}\cdot \vec{S}_{i+1})+J_{1}(\vec{S}_{i}\cdot \vec{S}_{i-1})
\label{eq:sc1_b}
\end{equation}
if $i\in c$;
\begin{equation} 
\begin{split}
 O^{SC}_{i,1}=&J_{2}(\vec{S}_{i}\cdot \vec{S}_{i+1})+J_{1}(\vec{S}_{i}\cdot \vec{S}_{i-1})\\
& +J_{3} (\vec{S}_{i}\cdot \vec{S}_{i-2})
\end{split}
\label{eq:sc1_c}
\end{equation}

\textit{\underline{Second order:}}

if $i\in a$;
\begin{equation} 
 \begin{split}
 O^{SC}_{i,2}=&J_{1}J_{2} (\vec{S}_{i}\cdot \vec{S}_{i+1})(\vec{S}_{i}\cdot \vec{S}_{i-1})+J_{1}J_{3} (\vec{S}_{i}\cdot \vec{S}_{i+1})\\
 &(\vec{S}_{i}\cdot \vec{S}_{i+2})
+J_{2}J_{3} (\vec{S}_{i}\cdot \vec{S}_{i-1})(\vec{S}_{i}\cdot \vec{S}_{i+2})
\end{split}
\label{eq:sc2_a}
\end{equation}
if $i\in b$;
\begin{equation} 
O^{SC}_{i,2}=J_{1}J_{1}(\vec{S}_{i}\cdot \vec{S}_{i+1})(\vec{S}_{i}\cdot \vec{S}_{i-1})
\label{eq:sc2_b}
\end{equation}

if $i\in c$;
\begin{equation} 
 \begin{split}
 O^{SC}_{i,2}=&J_{1}J_{2} (\vec{S}_{i}\cdot \vec{S}_{i+1})(\vec{S}_{i}\cdot \vec{S}_{i-1})+J_{1}J_{3} (\vec{S}_{i}\cdot \vec{S}_{i-1})\\
 &(\vec{S}_{i}\cdot \vec{S}_{i-2})
+J_{2}J_{3} (\vec{S}_{i}\cdot \vec{S}_{i+1})(\vec{S}_{i}\cdot \vec{S}_{i-2})
\end{split}
\label{eq:sc2_c}
\end{equation}

\section{Results}
\label{results}
Within the UCL approximation in the KH formalism, the RIXS spectra are mapped to a set of correlation functions in the SC and NSC channels, as previously described. We present the responses up to the second-order corrections of the UCL approximation. RIXS intensity is calculated by computing individual multi-spin correlation functions in the SC and NSC channels using the F-RGF technique~\cite{frgf-aes} on 18 sites containing six trimer unit cells and on 24 sites using Lanczos-based diagonalization \cite{lanczos} with eight trimer unit cells for $J_1 =1.0$, $\alpha=0.18$, and  $\beta=0.18$. We employ periodic boundary conditions in both cases. The real-space correlation functions are Fourier transformed to compute the momentum-dependent contributions to the RIXS intensity for all orders discussed in the previous section. The results of both methods are identical within numerical resolution for the same system size ($N=18$). We present the Lanczos results on $N=24$.  
\begin{figure*}[t]
	\centering
	\includegraphics[width=\linewidth]{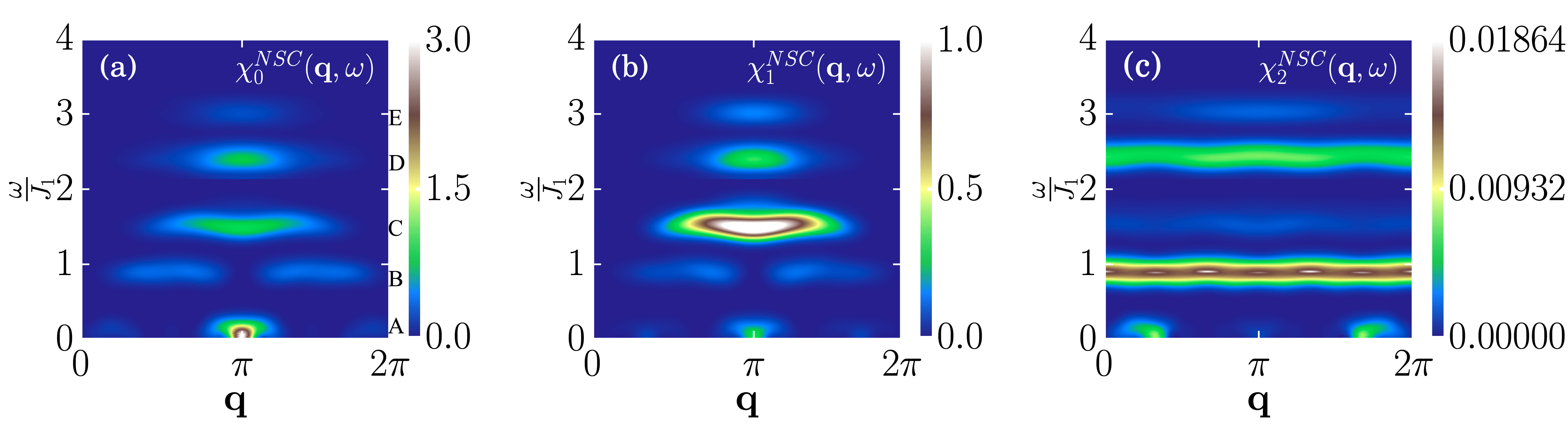}
	\caption{This figure shows the RIXS spectra in the non-spin conserving channel calculated for the parameters $J_{1}=1.0$ \& $J_{2}=0.18 J_{1}$ on $N=24$ sites. Panels (a), (b), and (c) display the zeroth order, first order, and second order terms, respectively, in the UCL expansion. To enhance the visibility of features beyond the energy threshold of  \textcolor{black}{$\omega = 2.135J_1$}, we have multiplied the RIXS intensity data by a factor of  500, 100, and 100 for panel (a), panel (b), and panel (c), respectively. \textcolor{black}{The low energy ($\omega <0.5J_1$) excitation data of panel (c) is multiplied by 10  for better visibility}}
\label{fig:spetra_nsc}
\end{figure*}
\subsection{Non-spin conserving channel}
The zeroth-order term within the NSC channel corresponds to the conventional dynamical susceptibility typically probed in inelastic neutron scattering experiments. In Fig.~\ref{fig:spetra_nsc} (a), the RIXS intensity is calculated using equations \eqref{eq:rixs_nsc} and \eqref{eq:nsc0}. This plot reveals five distinct features. The lowest feature, `A', is a gapless excitation extending up to $\omega/J_1\sim0.3$. This is followed by four distinct gapped excitations centered around  $\omega/J_1=0.9$, $\omega/J_1=1.5$, $\omega/J_1=2.4$, and $\omega/J_1=3.0$ labeled by B, C, D, and E respectively. To understand these features, we begin by analyzing the spectrum of the isolated trimer Hamiltonian. The trimer Hamiltonian is as follows:
\begin{figure}[]
    \centering
    \includegraphics[width=\linewidth]{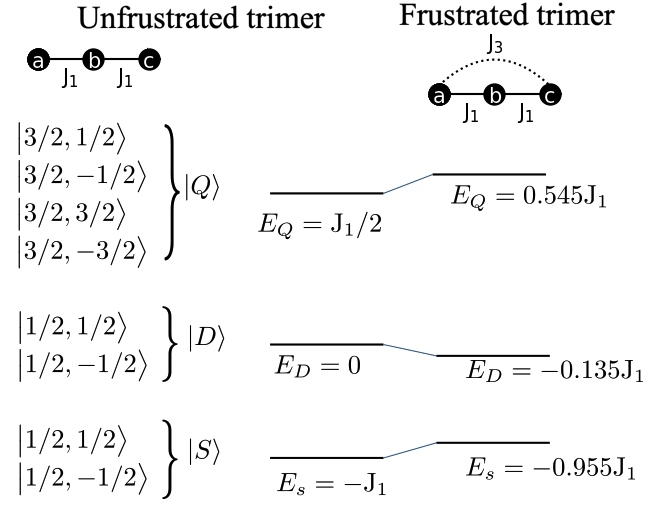}
    \caption{The level spectrum of a single unfrustrated and frustrated trimer: In the top left (right), we show the schematic of the unfrustrated (frustrated) trimer, respectively. The total spin quantum number $S^T$ and magnetic quantum number $S_{z}^{T}$ for all eigenstates \textit{which are identical for the unfrustrated and frustrated trimers} are provided below. The labels of spinon $|S\rangle$, doublon $|D\rangle$, and quarton $|Q\rangle$ are shown next to the eigenstates. We show the change in the energy levels up on introducing the frustrating interaction $J_3$. The eigenvalues of the unfrustrated and the frustrated trimers are shown in the middle and the right columns, respectively.}
    \label{fig:one_trimer}
\end{figure}

\begin{equation}
H_{Trimer}=J_{1}\vec{S}_{a}\cdot \vec{S}_{b}+J_{1}\vec{S}_{b} \cdot \vec{S}_{c}+J_{3}\vec{S}_{a} \cdot \vec{S}_{c}\nonumber
\end{equation}

For the \textit{unfrustrated} trimer we set $J_3=0$ in $H_{trimer}$. The three exact eigenvalues are $E_S=-J_{1}$, $E_D=0$ and $E_Q=0.5J_1$. The eigenstates in the $|S^{T}, S^{T}_{z}\rangle$ representation are shown in Fig.~\ref{fig:one_trimer} below the schematic of the unfrustrated trimer, where $S^{T}$ and $S^{T}_{z}$ are respectively, the total spin quantum number and total magnetic quantum number. The corresponding eigenvalues are shown in the middle column. The nomenclature of the states was introduced in literature\cite{Bera2022}. The lowest energy state is doubly degenerate, with the energy of $E_S = -J_{1}$ and labeled by $\ket{S}$. The next excited state, also doubly degenerate, has an energy of $E_{D}=0$ with label $\ket{D}$. While the $\ket{S}$ and $\ket{D}$ have the same $S^{T}$ and $S^{T}_z$, they differ in spatial character of the wavefunctions. From analyzing the eigenstates, we find that the states belonging to $\ket{S}$  with  $S^{T}_z=1/2$  and  $S^{T}_z=-1/2$ correspond to $\frac{1}{\sqrt{3}}(\uparrow\Bar{bc}-\Bar{ab}\uparrow)$ and $\frac{1}{\sqrt{3}}(\Bar{ab}\downarrow-\downarrow\Bar{bc})$ respectively, where $\Bar{ij}\equiv\frac{1}{\sqrt{2}}(|\uparrow_i\downarrow_j\rangle-\downarrow_i\uparrow_j\rangle)$, a singlet between spins at sites $i$ and $j$. Similarly the two states belonging to $\ket{D}$ with $S^{T}_z=1/2$  and $S^{T}_z=-1/2$ respectively correspond to \texttoptiebar{a$\uparrow$c} and \texttoptiebar{a$\downarrow$c}, where a singlet is formed between the `a' and `c' sites of the isolated trimer. 
The subsequent excited state exhibits four-fold degeneracy with the energy of $E_{Q} = 0.5J_{1}$ and is denoted by $\ket{Q}$. The $S_z^T=\pm 3/2$ states belonging to $\ket{Q}$ correspond to fully polarized states (up/down) for all spins of the trimer. The $S_z^T=+1/2$ and $S_z^T=-1/2$
states correspond to $\sigma=\uparrow$ and $\sigma=\downarrow$ respectively in the three spin state: $\frac{1}{\sqrt{6}}(\source{a}\target{b}\drawarrows\sigma +\source{a}\sigma\target{c}\drawarrows +\sigma\source{b}\target{c}\drawarrows~)$. Here $\source{i}\target{j}\drawarrows\equiv \frac{1}{\sqrt{2}}(|\uparrow_i\downarrow_j\rangle+|\downarrow_i\uparrow_j\rangle)$ is a triplet between spins at sites $i$ and $j$.
We now consider the \textit{frustrated} trimer. Up on switching on the $J_3=0.18J_1$ in $H_{trimer}$, we find the exact eigenvalues are altered and now occur at $E_S = -J_{1}+J_{3}$, $E_D =-\frac{3J_3}{4}$ and $E_Q=0.5 J_{1}+\frac{J_{3}}{4}$. These altered eigenvalues are shown in the right column in Fig.~\ref{fig:one_trimer}. It is clear the frustrating interaction $J_3$ raises the energy of the degenerate $\ket{S}$  and $\ket{Q}$ states while lowering the energy of the degenerate $\ket{D}$ states. Importantly, there are no level crossings or degeneracy lifting due to frustration, and hence, the unfrustrated state labels are carried over to frustrated cases. 

\begin{figure*}[t]
    \centering
    \includegraphics[width=0.8\linewidth]{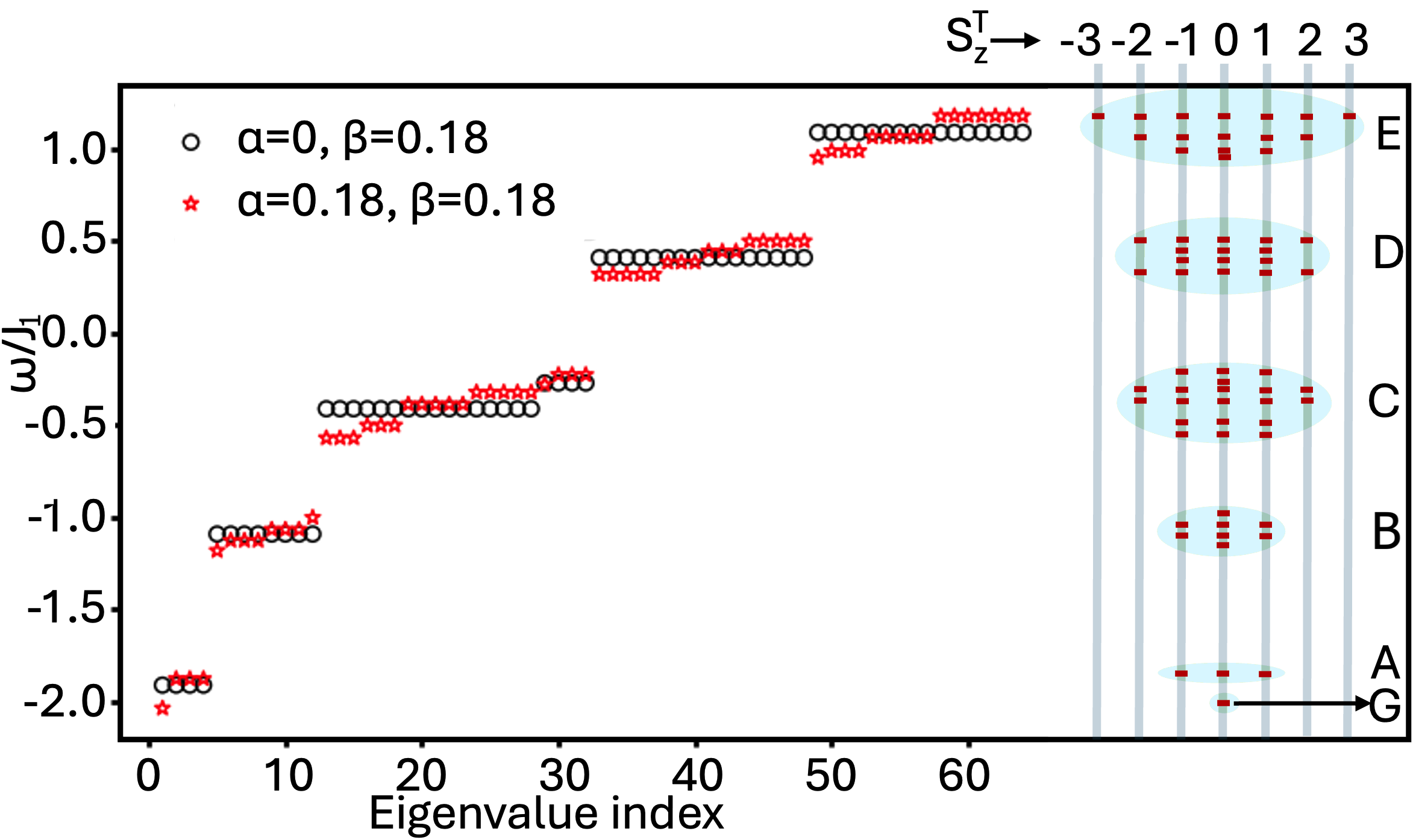}
    \caption{ In the left we show the eigenvalues for two decoupled (black circles) and coupled (red stars) trimers. On the right, we resolve the eigenvalues (red ellipses) in terms of the contributions from different $S_z^{T}$, the total $S_z$ of the basis components. The ground state is antiferromagnetic and is in the $S_z^{T}=0$ sector. The gray vertical lines denote the various $S_z^{T}$ values. The `A' to `E' refers to the excitations, and the blue shaded region is a guide to the eye, depicting the spread in the spectra when the two trimers are coupled, as compared to the uncoupled case. The energy locations of the `A' to `E'  features in the NSC channel in Fig.~\ref{fig:spetra_nsc}  are reported by setting the ground state energy to zero.}
    \label{fig:two_trimer_spec}
\end{figure*}

In previous literature \cite{Bera2022}, these three energy levels have been argued to be the dominant contribution excitations seen in INS-measured dynamical spin susceptibility for single spin-flip excitation. In this albeit simple but illuminating argument, the single spin-flip in a trimer `a' or `c' sites upon coupling to other trimers was expected to broaden into a gapless fractionalized spinon continuum as would happen in an antiferromagnetic ground state. Thus setting $E_S=0$ the energy levels $E_D$,  and $E_Q$ are determined to be $\omega/J_1=0.82$ and $\omega/J_1=1.5$.  The locations of these features agree with the centroids of the `B' and the `C' features in the single spin-flip INS excitation  \cite{Bera2022} and to the zeroth order of the NSC RIXS spectra shown in   Fig.~\ref{fig:spetra_nsc} (a). These excitations were denoted as  fractionalized spinon ($|S\rangle$), doublon ($|D\rangle$), and quarton ($|Q\rangle$) modes. While the frustrating exchange $J_3$ is small, we find that it is important to include it in the analysis (unlike the $J_1-J_2$ trimer spin chain\cite{Cheng2022-sum-rule}) for correctly predicting the energy location of both the INS and the RIXS excitation spectra. 

However, the high energy excitations in Fig.~\ref{fig:spetra_nsc} in the lowest order in the RIXS NSC channel,  `D' and `E', cannot be captured with a single spin trimer. Moreover, as seen from the connectivity in the RIXS operators, the first and second order in the UCL expansion (Fig.~\ref{fig:operator_coonec}) couple two spin-trimers. We, therefore, provide a more accurate analysis by considering two spin trimers and show that the excitations of SC and NSC channels can be understood in a comprehensive manner.

We consider two isolated frustrated trimers with three energy levels: $E_{S}$, $E_{D}$, and $E_{Q}$ individually. The Hilbert space is spanned by nine states constructed from the single trimer states $|S\rangle$, $|D\rangle$ and $|Q\rangle$. It easy to see that there are nine eigenvalues  $2E_{S}$, $2E_{D}$, $2E_{Q}$, with doubly degenerate $E_S + E_{D}$, $2E_{D}+E_Q$ and $E_{S}+E_Q$, yielding six distinct energy levels. In Fig.~\ref{fig:two_trimer_spec}, we show these six eigenvalues of two with circles. \textcolor{black}{These eigenvalues in increasing order of magnitude are  $2E_{S}=-1.91 J_1$,  $E_S + E_{D}=-1.09 J_1$,$E_S + E_{Q}=-0.41 J_1$, $2E_{D}=-0.27 J_1$, $E_D + E_{Q}=0.41 J_1$ and $2E_Q =1.09 J_1$.}

\begin{figure*}[t]
	\centering
	\includegraphics[width=0.8\linewidth]{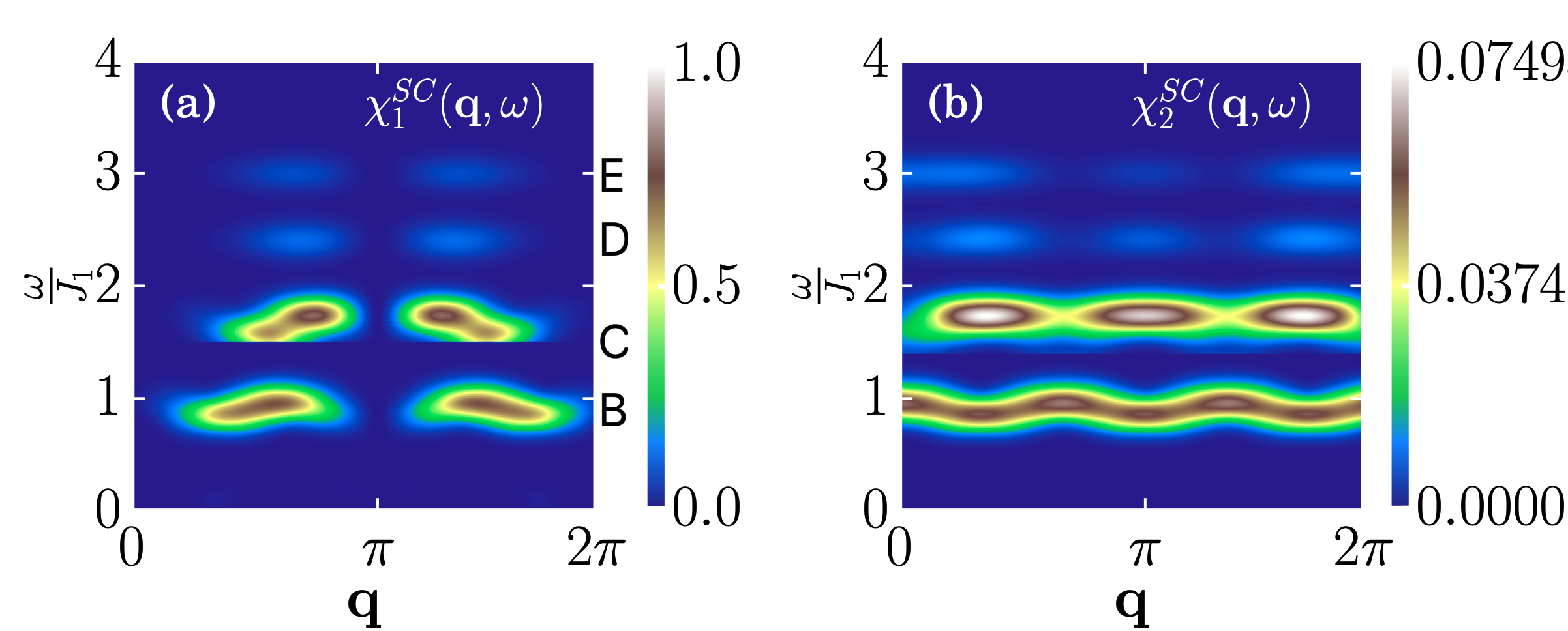}
	\caption{Panels (a) and (b) illustrate the RIXS spectra in the spin-conserving (SC) channel, calculated using parameters $J_{1}=1.0$, and $J_{2}=J_3=0.18 J_{1}$ for the first and second order terms, respectively on $N=24$ sites. To improve the visibility of features beyond the energy threshold of \textcolor{black}{$\omega = 1.35J_1$}, we have amplified the RIXS intensity data by a factor of 100 for values exceeding this threshold.}
	\label{fig:spetra_sc}
\end{figure*}
The stars denote the eigenvalues for the coupled timers  ($\alpha =0.18$, $\beta=0.18$). When the two frustrated trimers are coupled, the eigenvalue spectrum is altered non-trivially. Firstly, we identify a $S^T_z=0$ eigenvalue with the lowest energy as the ground state labeled as `G' with $E_{G}=-2.03539J_1$. This is followed by a state with energy close to $2E_{S}$ labeled as `A.' Since in the spinon state of an isolated frustrated trimer, there was one spin 1/2 and a singlet, a state dominantly made out of two spinon states on individual trimers can have $S_Z^T=-1,0,1$ as indicated on the right of the feature `A' in Fig.~\ref{fig:two_trimer_spec}. 
Similarly, the `B' state is centered around $E_S+E_D=-1.09J_1$, primarily composed of a doublon and a spinon. The result of coupling the trimers further results in the overlap of the energy levels $2E_{D}=-0.27J_1$ and $E_{S}+E_Q=-0.41J_1$ resulting in a single feature 'C'. The next two energy states are broadened around $E_{D}+E_Q=0.41J_1$ marked as `D' and around $2E_Q=1.09J_1$ marked as `E'. 
We note that at least two timers are needed to capture the  $S^T_z=0$ ground state. Also, we find that only five well-separated features remain apart from the ground state. \textcolor{black}{Measuring the separation of the these from the ground state, we find the energy locations of the five features to be $\Delta\omega_A=0.125J_1$,  $\Delta\omega_B=0.945J_1$,  $\Delta\omega_C=1.625J_1$, $\Delta\omega_D=2.445J_1$ and  $\Delta\omega_E=3.125J_1$, in close agreement with the feature locations in Fig.~\ref{fig:spetra_nsc}
(a). In addition, we identify that the `D' and the `E' features arise dominantly from a `doublon and a quarton', and two quarton excitations,  respectively. We note that $\Delta\omega_A=0.125J_1$ is expected to go to zero with increasing system size. We also see that the energy centroids of the `B' and `C' features are approximately equal to $E_D$ and $E_Q$, respectively, since $E_S<<E_D<<E_Q$. While crude, this explains why the single frustrated trimer eigenvalues could label the INS data as discussed earlier. However, our analysis shows that there is a contribution of spinons to the  B feature and that the C feature is an admixture of the spinon, doublon and quarton states.}

The first order correction in the NSC channel are computed using Eq.~\eqref{eq:rixs_nsc} and Eq.~\eqref{eq:nsc1_a},\eqref{eq:nsc1_b},\eqref{eq:nsc1_c} and shown in Fig \ref{fig:spetra_nsc} (b). Feature locations in energy in this order are the same as for the zeroth order, except that the spectral weight of the quarton modes is increased while spinon and doublon modes are softened. We have shown the second-order correction in the NSC channel in Fig.~\ref{fig:spetra_nsc} (c) and calculated by using Eq.~\eqref{eq:rixs_nsc} and Eq.~\eqref{eq:nsc2_a},\eqref{eq:nsc2_b},\eqref{eq:nsc2_c}. This demonstrates a similarity to the zeroth-order correction, highlighting five different features. At this order, spinon stiffens at the edge of the Brillouin zone, and quarton modes and feature `D' appear throughout the whole Brillouin zone. The higher-order operators include the possibility of double spin-flip excitations at and around the core-hole creation site. This results in renormalization of the single-spin flip excitation at the core-hole site. We note that the NSC channel operators contain a single spin-flip operator multiplying \textit{scalar} two spin-flip operators and hence only connect states differing by one unit of spin angular momentum. In Fig.~\ref{fig:two_trimer_spec} on the right side, we provide the $S^T_z$ of the ground state and the five excitations for the coupled frustrated trimers. Thus, the excitations in the NSC channel occur between states separated by $\Delta S^T_z=\pm 1$. This also implies that while the features renormalize due to two spin-flip excitations, they create only those excitation energies as in the zeroth order. Thus, the energy locations of the features remain the same in higher orders. 

\subsection{Spin conserving channel}
We now consider the SC channel RIXS spectra. As seen from Appendix \ref{appendix} Eq.~\eqref{SC-L}, the zeroth order in the SC channel contains contributions of charge correlations. Thus, the lowest-order spin correlation term in the SC channel comes from the first-order contribution of the UCL expansion. The first-order term in the SC channel involves a double spin flip that conserve\sout{d}\uk{s} $S^{T}_{z}$. Fig.~\ref{fig:spetra_sc} (a) shows the first order correction in the SC channel computed by using Eq.~\eqref{eq:spectra_sc} and \eqref{eq:sc1_a},\eqref{eq:sc1_b},\eqref{eq:sc1_c}. A key characteristic of this order is that the spectral weight vanishes at $q=0$ and $\pi$. At $q=0$, matrix element $\sum_{f}\langle f| \frac{1}{\sqrt{N}}\sum_{i=1}^{N} O_{i,1}^{SC} |g\rangle$ vanishes because operator $O_{i,1}^{SC}$ at q=0 commutes with the Hamiltonian resulting in vanishing RIXS intensity. At $q=\pi$, the operator can be written as $\sum_{i=1}^{N}e^{\iota q r_{i}} O_{i,1}^{SC}=\sum_{i\in even}O_{i,1}^{SC}-\sum_{i\in odd}O_{i,1}^{SC}$ and this leads to the vanishing of matrix element which leads to the vanishing intensity. 
\textcolor{black}{The excitation created by two spin flips is fractionalized and gives five distinct features labeled A, B, C, D, and E, as in the NSC channel.} Since the operators evaluated in the SC channel are scalars, the only allowed transitions conserve $S_z^T=0$ between the ground and excited states. This selection rule constrains the number of states in the excitation manifold to have non-zero matrix elements with the $S_z^T=0$ ground state, as is apparent from the reduced number of possible states excitation states connecting to the ground state as seen on the right side of Fig.~  \ref{fig:two_trimer_spec}. We note for a single trimer, due to the selection rule, the contribution to the SC channel is identically zero for the 'A' feature. We find a highly suppressed but finite $(\sim10^{-2})$ contribution for coupled two trimers at `A.' It is, however, not visible in the color plot in Fig.~\ref{fig:spetra_sc} (a).

Most of the spectral weight in this order appears in `B' and `C' features. Additionally, high-energy features `D' and `E', with highly suppressed contributions, also appear in the SC channel. The second-order correction in the SC channel is four spin flip terms. It is calculated by using Eq.~\eqref{eq:sc2_a},\eqref{eq:sc2_b},\eqref{eq:sc2_c} in Eq.~\eqref{eq:spectra_sc} and  depicted in Fig.~\ref{fig:spetra_sc} (b). In contrast to the first-order correction in the SC channel, the' B' feature hardens in second-order correction and appears through the whole Brillouin zone, while `C', D, and E modes exhibit the three sub-structures.

\section{Conclusion}
\label{concl}
In this study, we have investigated the Resonant Inelastic X-ray Scattering (RIXS) cross-section for a quasi-1D frustrated antiferromagnetic Heisenberg model. The model parameters are chosen to mimic the quasi-1D frustrated spin timers that are weakly coupled in  Na$_{2}$Cu$_{3}$Ge$_{4}$O$_{12}$. We have computed the RIXS spectra for the NSC and SC channels within the framework of the UCL expansion of the KH formalism, extending up to the second order. In this paper, we have focused only on the RIXS-induced many-body excitations. 

The intra-trimer interaction in the spin-$\frac{1}{2}$ trimer chain in Na$_2$Cu$_3$Ge$_4$O$_{12}$ is reported to be $J_1 = 20.21$ meV in recent experimental study. With this interaction strength, the lowest three features are dominantly made of fractionalized spinon, doublon, and quarton modes at approximately $\omega = 0.3 J_1 = 6.063$ meV, $\omega = 0.9 J_1 = 18.199$ meV, and $\omega = 1.5 J_1 = 30$ meV, respectively, which is consistent with the recent work. 

Our analysis further shows that the lowest feature eventually forms a gapless spinon continuum. The next two features, `B' and `C', have an admixture of `spinon and doublon', and `spinon, doublon and quarton' modes, respectively, improving on the previous reports. Moreover, the study predicts two new and significant features, D and E, which are expected to emerge at approximately $\omega = 2.4 J_1 = 48$ meV and $\omega = 3.0 J_1 = 60$ meV, respectively. These are made up of a combination of `doublon and quarton', and `two quarton' modes respectively. 
We have further shown that the SC channels RIXS show spectral features at the same energy as in the NSC channel but are significantly suppressed in comparison. We also predict the existence of band-like low-energy excitations, spreading over the entire Brillouin zone in the second order of the NSC and SC channels. The energy range of these features is easily accessible to $K$ and $L$ edge RIXS resolution. This, therefore, makes RIXS an ideal probe to explore these higher energy excitations.

 Understanding the fractionalization of multi-spin correlations is an actively pursued area and holds the key to understanding the nature of collective excitations in correlated systems. Our results highlight the potential of RIXS to understand fractionalized excitations in quantum materials.

\section{Acknowledgements:}
All computations were performed in the NOETHER, VIRGO high-performance clusters at NISER. AM acknowledges helpful discussions with V. Ravi Chandra at NISER.

\appendix
\begin{center}
	\textbf{APPENDIX}
\end{center}

\section{Ultra-short core-hole lifetime expansion of Kramers-Heisenberg formalism}
\label{appendix}
Here, we provide ultra-short core-hole lifetime expansion of Kramers-Heisenberg formalism. The RIXS intensity depends on the  Kramers-Heisenberg scattering amplitude ($A_{fg}$) between the initial state of the system $|g\rangle$, which is the ground state, and a final state $|f\rangle$, following the decay of the core-hole. This evolution from the $|g\rangle$ to the final state $|f\rangle$occurs through the states of the intermediate state Hamiltonian $H_{int}$ which comprises of the unperturbed Hamiltonian $H_0$ before photo-excitation and perturbative corrections ($H^\prime$), from the core hole induced by X-ray photons. Similar to the initials state, the possible final states $|f\rangle$ also \textit{do not} contain the core-hole. Unlike these states, however, the core hole is present during the evolution between $|i\rangle$ and  $|f\rangle$ by intermediate $H_{int}$, whose eigenproblem is defined by $H_{int}|n\rangle=E_n|n\rangle$. These definitions will be useful below. The final cross-section is summed over all possible final states as described below.

\begin{equation}
\frac{d^2\sigma}{d\Omega d\omega} \propto  \sum_{f} |A_{fg}|^2 \delta(E_f-E_g-\omega)
\label{HK_1}
\end{equation}
Here
\begin{equation}
A_{fg} = \omega_{res}\sum_n \frac{\langle f | \hat{D} | n \rangle\langle n | \hat{D} | g\rangle}{\omega_{in} - E_n +i\Gamma}
\label{A_fg}
\end{equation}
The Kramers-Heisenberg amplitude $A_{fg}$ is defined in Eq.~\eqref{A_fg}, representing the transition amplitude between the initial state $|g\rangle$ with energy $E_g$ (used as reference energy: $E_g=0$) and the final state $|f\rangle$. The initial state $|g\rangle$ is photo-excited to an intermediate state described by the dipole operator $\hat{D}$, and $\Gamma$ is a broadening factor stemming from the lifetime of the core hole. The dipole operator for the  $K$-edge and $L$-edge RIXS cases is well documented. The main approach to simplifying $A_{fg}$ is ultra-short core-hole lifetime, implying $\Gamma$ as the largest scale in the denominator of  $A_{fg}$. The detuning of the incoming photon energy is given by $\omega_{in} = \omega_{in}^0 - \omega_{res}$, where $\omega_{res}$ is the energy of the target $K$ or $L$ edge core-level. A part of the incoming photon energy is absorbed during the relaxation process of the core hole so that the energy of the outgoing photon is smaller than $\omega_{in}^0$. The $\delta$-function in Eq \eqref{HK_1}, ensures that the energy of the outgoing photon and core-hole relaxation respect energy conservation. 

Now, by identifying a detuning parameter $\Delta=\omega_{in}-i\Gamma$ in Eq.~\eqref{HK_1}, we expand $A_{fg}$ as, $\frac{\omega_{res}}{\Delta}\sum_{l=0}^\infty\langle f|DH_{int}^lD|g\rangle/\Delta^{(l)}$.
Also, for $H_0$, the Heisenberg Hamiltonian, and with the usual assumption of local core-hole\cite{PhysRevB.77.134428}, $H_{int}$ becomes $H_0+H^\prime$, where  $H^\prime=\eta\sum_{ij} h_ih_i^\dagger J_{ij}S_i.S_j$. Here $h_i$ ($h_i^\dagger$)  creates (destroys) a core hole at site $i$. In the cases of exact resonance $\omega_{in}=0$, we have:

\begin{equation}
A_{fg} = \frac{\omega_{res}}{(i\Gamma)}\sum_{l=0}^\infty\langle f|D\frac{(H_{0}+H^\prime)^l}{i\Gamma^l}D|g\rangle
\label{A_expan}
\end{equation}
The dipole operator $D=D_0+D_0^\dagger$ , where
$\hat{D}_{0}(\epsilon,\epsilon')=\sum_{i\sigma}(n_{i,\sigma} r_{\epsilon'}e^{iq_{out} R_{i}}d^\dagger_i p_i + r_{\epsilon}e^{-iq_{in}R_{i}}p^\dagger_i d_i)
$ for spin-conserving $L$-edge scattering and
 $\hat{D}_{0}(\epsilon,\epsilon')= \sum_i (S^{x}_i r_{\epsilon'}e^{iq_{out}R_{i}}d^\dagger_i p_i + r_{\epsilon}e^{-iq_{in} R_{i}}p^\dagger_i d_i)
\label{dipol_NSC}$ for the spin non-conserving $L$-edge scattering. In the later expression, $S_i^x$ can induce a single spin flip (leading to the spin-conserving scattering), $r_{\epsilon}$ and $r_{\epsilon'}$ denote the amplitude of the incoming and outgoing photons with polarization $\epsilon$ and $\epsilon'$, respectively. They give rise to the polarization-dependent matrix elements for SC and  NSC channels. This polarization dependence is different for the SC and NSC channels. Using the fact that $[D,H_0]=0$ and the ground state energy $E_g$ is defined to be zero, RIXS intensity  for the SC and NSC channels can be expressed as,
\begin{widetext}
    \begin{align}
&I^{NSC}(q,\omega)\propto \big( \frac{1}{\Gamma^2}\sum_f \Big|\langle f|\frac{1}{\sqrt{N}}\sum_i e^{iq R_{i}}S_i^{x}|g\rangle\Big|^2
+ \frac{1}{\Gamma^4}\sum_f \Big|\langle f|\frac{1}{\sqrt{N}}\sum_{i,j} e^{iqR}_{i}J_{i,j}S_i^{x}(\hat{S}_i\cdot  \hat{S}_j) |g\rangle\Big|^2 &\nonumber\\
&+\frac{1}{\Gamma^6}\sum_f \Big|\langle f|\frac{1}{\sqrt{N}}\sum_{i,j,k} e^{iqR_{i}}J_{i,j}J_{i,k}S_i^{x}(\hat{S}_i\cdot  \hat{S}_j)(\hat{S}_i\cdot  \hat{S}_k) |g\rangle\Big|^2 + \cdot \cdot\cdot\Big)\delta\left(E_f-E_g-\omega\right)\nonumber\\
&=\sum_{l}\chi_{l}^{NSC}(q,\omega)
\label{NSC-L}
\end{align}

\end{widetext}

\begin{widetext}

\begin{align}
&I^{SC}(q,\omega)\propto \big( \frac{1}{\Gamma^2}\sum_f \Big|\langle f|\frac{1}{\sqrt{N}}\sum_i e^{iqR_{i}}n_{i,\sigma}|g\rangle\Big|^2
+ \frac{1}{\Gamma^4}\sum_f \Big|\langle f|\frac{1}{\sqrt{N}}\sum_{i,j} e^{iqR_{i}}J_{i,j}\hat{S}_i\cdot  \hat{S}_j |g\rangle\Big|^2 &\nonumber\\
&+ \frac{1}{\Gamma^6}\sum_f \Big|\langle f|\frac{1}{\sqrt{N}}\sum_{i,j,k} e^{iqR_{i}}J_{i,j}J_{i,k}(\hat{S}_i\cdot  \hat{S}_j) (\hat{S}_i\cdot  \hat{S}_k)|g\rangle\Big|^2 + \cdot \cdot \cdot\textbf{}\Big)\delta\left(E_f-E_g-\omega\right)\nonumber \\
&=\sum_{l}\chi_{l}^{SC}(q,\omega)
\label{SC-L}
\end{align}

\end{widetext}
In the NSC channel, $O((1/\Gamma)^2)$  term is the single spin-flip spin excitation scattering, and the $O((1/\Gamma)^4)$ term is a combination single spin-flip at the site where the core-hole is created and a double flip at $i$ and $j$ site. $O((1/\Gamma)^6)$ combination single spin-flip at the site where the core-hole is created and a double flip at $j$ and $k$.
For the spin-conserving channel, the zeroth order term contributes only to elastic scattering.$O((1/\Gamma)^4)$ term involves double spin flip at $i$ and $j$. In contrast, $O((1/\Gamma)^6)$ is the combination of a double spin flip at $i$ and $k$ followed by another double spin flip at $i$ and $j$. 


\bibliography{bibliography.bib}
\end{document}